\documentclass[conference]{IEEEtran}
\usepackage{indentfirst}
\usepackage{amsmath}
\usepackage{amsfonts}
\usepackage{amssymb}
\usepackage{cite}
\usepackage{times}
\usepackage[dvips]{graphicx}
\usepackage{hhline}
\usepackage{color}
\usepackage{multirow}
\usepackage{graphicx}
\usepackage{epstopdf}
\usepackage[acronym,shortcuts]{glossaries}
\usepackage[T1]{fontenc}
\usepackage[utf8]{inputenc}
\usepackage[english]{babel}
\usepackage{balance}
\usepackage{amsmath, mathtools}
\usepackage{amsfonts,amsthm,bm}
\usepackage{amssymb}
\usepackage{dsfont}
\usepackage{graphicx}
\usepackage{mathtools}
\usepackage{color, colortbl}
\usepackage{soul}
\usepackage[acronym,shortcuts]{glossaries}
\usepackage{tikz}
\newcommand{\myempth}[1]{"#1"}
\IEEEoverridecommandlockouts

\newacronym{ar}{AR}{achievable rate}
\newacronym{awgn}{AWGN}{additive white Gaussian noise}
\newacronym{bd}{BD}{block diagonalization}
\newacronym{csi}{CSI}{channel state information}
\newacronym{df}{DF}{decode-and-forward}
\newacronym{dft}{DFT}{discrete Fourier transform}
\newacronym{b5g}{B5G}{fifth generation}
\newacronym{gnb}{gNB}{new-generation node base}
\newacronym{ga}{GA}{genetic algorithm}
\newacronym{los}{LoS}{line-of-sight}
\newacronym{mimo}{MIMO}{multiple-input multiple-output}
\newacronym{miso}{MISO}{multiple-input single-output}
\newacronym{siso}{SISO}{single-input single-output}
\newacronym{mu-mimo}{MU-MIMO}{multi-user multiple-input multiple-output}
\newacronym{su-mimo}{SU-MIMO}{single-user multiple-input multiple-output}
\newacronym{snr}{SNR}{signal-to-noise ratio}
\newacronym{svd}{SVD}{singular value decomposition}
\newacronym{ue}{UE}{user equipment}
\newacronym{af}{AF}{amplify-and-forward}
\newacronym{hd}{HD}{half-duplex}
\newacronym{fd}{FD}{full-duplex}
\newacronym{fic}{FIC}{fast iterative configuration}
\newacronym{6g}{6G}{sixth generation}
\newacronym{mmwave}{mmWave}{millimetre-wave}
\newacronym{2d}{2D}{two-dimensional}
\newacronym{rf}{RF}{radio-frequency}
\newacronym{nlos}{NLOS}{non-line-of-sight}
\newacronym{aoa}{AoA}{angle of arrival}
\newacronym{aod}{AoD}{angle of departure}
\newacronym{ula}{ULA}{uniform linear array}
\newacronym{upa}{UPA}{uniform planar array}
\newacronym{omp}{OMP}{orthogonal matching pursuit}
\newacronym{cs}{CS}{compressive sensing}
\newacronym{ris}{RIS}{Reconfigurable intelligent surface}

\DeclareMathOperator{\diag}{diag}

\newcommand{\bi}{\begin{itemize}}
\newcommand{\ei}{\end{itemize}}

\newcommand{\be}{\begin{IEEEeqnarray}}
\newcommand{\ee}{\end{IEEEeqnarray}}

\newcommand{\commentLB}[1]{}

\begin{document}
	
	\title{Fast Iterative Configuration of Reconfigurable Intelligent Surfaces in mmWave Systems}

	\author{ 
		Anna V. Guglielmi$^{1}$
		 and Stefano Tomasin$^{1, 2}$ \\
		
		$^{1}$ Dept.\ of Information Engineering, University of Padova, Italy \\
		$^{2}$ Consorzio Nazionale Interuniversitario di Telecomunicazioni, CNIT, Parma, Italy\\

		email: \{annavaleria.guglielmi, stefano.tomasin\}@unipd.it

\thanks{Project funded under the National Recovery and Resilience Plan (NRRP), Mission 4 Component 2 Investment 1.3 - Call for tender No. 341 of 15 marzo 2022 of Italian Ministry of University and Research funded by the European Union – NextGenerationEU Project code MUR: PE\_0000001 Concession Decree No. 1549 of 11/10/2022 adopted by the Italian Ministry of University and Research, CUP C93C22005250001 Project title RESearch and innovation on future Telecommunications systems and networks, to make Italy more smART (RESTART)}
%\vspace{-1.5cm}
}
	
	\date{}
	\maketitle
	\thispagestyle{empty}
	\pagestyle{empty}

\begin{abstract}
\acp{ris} are a promising solution to improve the coverage of cellular networks, thanks to their ability to steer impinging signals in desired directions. However, they introduce an overhead in the communication process since the optimal configuration of a \ac{ris} depends on the channels to and from the \ac{ris}, which must be estimated. In this paper, we propose a novel \ac{fic} protocol to determine the optimal \ac{ris} configuration that exploits the small number of paths of \ac{mmwave} channels and an adaptive choice of the explored \ac{ris} configurations. In particular, we split the elements of the \ac{ris} into a number of subsets equal to the number of channel taps. For each subset, then an iterative procedure finds at each iteration the optimal \ac{ris} configuration in a codebook exploring a two-dimensional grid of possible angles of arrival and departure of the path at the \ac{ris}. Over the iterations, the grid is made finer around the point identified in previous iterations.  Numerical results obtained using an urban channel model confirm that the proposed solution is fast and provides a configuration close to the optimal in a shorter time than other existing approaches. 
\end{abstract}

\begin{IEEEkeywords} 
	Reconfigurable intelligent surfaces (RISs), channel estimation, beamforming.
\end{IEEEkeywords}

\glsresetall

\section{Introduction}
\label{sec:intro}
\Acp{ris} are seen as a key enabling technology for beyond \ac{b5g} wireless systems, due to their ability to purposely shape wireless environments \cite{huang2019}. %A typical \ac{ris} generally consists of multiple controllable reflective elements that introduce a phase shift in the baseband equivalent wireless signal, and such shift can be adjusted  by means of lightweight electronics.  %\acp{ris} represent a cost-effective solution to the link blockage problem at high frequencies, e.g., in the millimeter-wave \ac{mmwave} bands, by establishing a controllable reflected link.  However, \acp{ris} need to be large, specifically when dealing with far-field communications so that they can compensate for the significant end-to-end path-loss caused by the multiplication of the individual path-losses of the transmitter-to-\ac{ris} and \ac{ris}-to-receiver channels.
%Furthermore, \acp{ris} have proven to be a multifaceted tool allowing improvements in localization, radio and radar mapping, security, energy efficiency, and electromagnetic field exposure \cite{strinati2021, strinati2021_2}.
%
An important challenge lies in the optimization of the phase-shifts, 
%For instance, the optimization can be targeted to steer the reflected signal in the direction of desired users as well as to optimize localization and sensing performance \cite{molisch2017}. 
since beam management in \ac{mmwave} communications is a key technique to overcome the severe free-space path loss and atmospheric absorption \cite{giordani2019, attaoui2022}. However, two main difficulties arise in \ac{ris} optimization: (i) the problem is non-convex, due to the unit modulus constraint; (ii) the impact of phase shifts on performance metrics such as the \ac{snr} or the achievable rate is typically coupled with other unknown system model parameters. 

%It is worth noting that the optimization can be done before or after performing the channel estimation. From the literature, it emerges that for data communication purposes, the optimization after channel estimation can be solved based on the estimated cascade channel response. Once the cascade channel estimation is available, the proper pair of precoder/combiner and phase shifts configuration are identified and communicated to the \ac{ris} via a side control link \cite{he2020, you2020}. Recently, alternative approaches have been proposed to mitigate the overhead due to cascade channel estimation and reduce the delay introduced by the \ac{ris} optimization, including beam training and exploiting the \ac{ris} diversity gain \cite{psomas2021, alexandropoulos2022, an2022}. Neverthless, the knowledge of the underlying channel is quite challenging to acquire: even in the case in which the channel model is linear and known, imprecise knowledge of the parameters of the channel, i.e., \ac{csi}, can significantly degrade the performance.  The optimization before channel estimation can be addressed with approaches such as maximizing recovery performance by means of \ac{dft} or Hadamard matrices, or exploiting external location information \cite{hu2020, rahal2022_2}. 

The majority of the existing works aim at optimizing the \ac{ris} elements individually for real-time communications, leading to computationally expensive solutions when the number of elements is large. It has been shown that for usual outdoor scenarios, hundreds or even thousands of \ac{ris} elements are needed to generate a virtual \ac{los} link as strong as a \ac{los} \cite{najafi2021}. Thus, most algorithms in the literature are not suitable for large \acp{ris}: due to their complexity, several solutions have been considered. To overcome this problem, the design of a codebook of phase-shift configurations in an offline fashion is proposed in \cite{najafi2021}, followed by the selection of the best codeword from the set in an online optimization fashion for the given channel realization. %In this way, the complexity of the online optimization and the corresponding channel estimation overhead scale with respect to the codebook size instead of the number of \ac{ris} elements. 
In \cite{zheng2020_1, zheng2020_2}, predefined phase-shift configurations based on the \ac{dft} matrix for channel estimation in a \ac{ris}-assisted system are employed. Nevertheless, \ac{dft}-based codebooks consist of the same number of phase-shift configurations as the number of \ac{ris} elements, which are too many for practical implementation. Consequently, in \cite{jamali2021} a quadratic codebook design is proposed, featuring a parameter for the control of the codebook size. It has been shown that under the far-field assumption, the overhead of beam training with exhaustive search can be significantly reduced
using small-size \ac{ris} phase shifts codebooks. A variable-width hierarchical phase-shift codebook suitable for both the near- and far-field of the \ac{ris} in \ac{mmwave} communication systems was proposed in \cite{alexandropoulos2022_2}. Instead, \cite{najafi2021, zheng2020_1, zheng2020_2, jamali2021, alexandropoulos2022_2} focus on continuous \ac{ris} phase shifts; however, discrete \ac{ris} phase shifts are to be preferred to reduce the implementation cost at the expense of increasing the complexity of the optimization problem. Recently, in \cite{rahal2022, rahal2022_3} a generic framework has been proposed, aiming at optimizing the end-to-end precoder controlled by \acp{ris} so that arbitrary beam patterns can be generated, given a predefined lookup table of \ac{ris} element-wise complex reflection coefficients. In \cite{ghanem2022} the \ac{ris} optimization is framed as two offline optimization problems towards a favorable trade-off between power consumption and codebook size. However, in \cite{rahal2022, rahal2022_3,ghanem2022, he2020, you2020}, the \ac{ris} configuration optimization is based on the knowledge of the channels to and from I, which is not easily obtainable. 
%response. Once the cascade channel estimation is available, the proper pair of precoder/combiner and phase shifts configuration are identified and communicated to the \ac{ris} via a side control link \cite{he2020, you2020}. 

Alternative methods have been proposed to mitigate the overhead due to cascade channel estimation and reduce the delay introduced by the \ac{ris} optimization, including beam training and exploiting the \ac{ris} diversity gain \cite{psomas2021, alexandropoulos2022, an2022}. Nevertheless, again obtaining an accurate knowledge of the underlying channel is quite challenging: even when the channel model is linear and known, the imprecise knowledge of the parameters of the channel, can significantly degrade the performance. To overcome this problem, the optimization can be done before channel estimation. However, only a few works address it and employ approaches such as maximizing recovery performance by means of \ac{dft} or Hadamard matrices, or exploiting external location information \cite{hu2020, rahal2022_2}, which however may not be accurate enough and becomes problematic in the presence of \ac{nlos} propagation conditions. % In particular, the localization-optimal phase profile is designed with the aim of minimizing the position error bound (PEB) assuming imperfect user location information \cite{hu2020} or prior knowledge of the user equipment location \cite{rahal2022_2}.

In this paper, we propose the novel \ac{fic} algorithm that configures the \ac{ris} elements for the uplink of a cellular system operating at \ac{mmwave} frequencies. We assume that the channels are unknown and we leverage the fact that at those frequencies the number of channel paths is very small. We also exploit the fact that the receiving base station (which can estimate the channel) is also controlling the \ac{ris}. Thus, we propose an adaptive iterative algorithm. \ac{fic} first partitions the \ac{ris} elements into subsets, where each subset is configured to reflect one path. For each subset, we apply an iterative algorithm, where at each iteration the source-\ac{ris}-destination channel is estimated for a set of \ac{ris} configurations, and the resulting achievable rate is computed. Each \ac{ris} configuration is associated with a couple of \ac{aoa} and \ac{aod} at the \ac{ris}, and the angles of the set are on a regularly-spaced two-dimensional (2D) angle grid. The configuration providing the highest achievable rate is selected and its \ac{aod}-\ac{aoa} couple becomes the center of a second smaller and finer grid for the next iteration. The configuration of the \ac{ris} for a given channel realization is thus iteratively adjusted towards the maximization of the achievable rate. 
It is worth noting that the proposed approach does not assume the complete knowledge of the channels from and to the \ac{ris}, and considers discrete phase-shift values.

%The peculiarities of the proposed \ac{fic} solution are a) the choice of the \ac{ris} configuration to be tested in the next iteration is guided by the achievable rate obtained by the configurations of the previous iteration, thus the process is an adaptive optimization; b) the procedure is tailored to a \ac{mmwave} scenario with few paths per channel.  

The rest of the paper is organized as follows. Section \ref{sec:sysmod} describes the system model. Section~\ref{FIC} describes the \ac{fic} algorithm and the design of the explored \ac{ris} configurations. Numerical results are presented and discussed in Section~\ref{sec:numres}. Lastly, Section~\ref{sec:conc} concludes the paper.

\section{System model}
\label{sec:sysmod}

\begin{figure}
 \centering
 \includegraphics[width=\columnwidth]{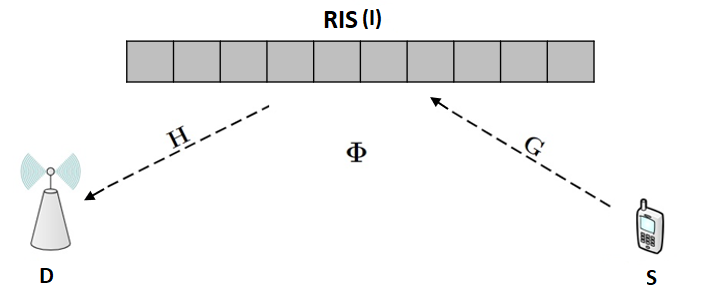} \vspace{-1cm}
  \caption{System model.}
  \label{fig:system_model}
\end{figure}

We consider a narrowband single-user \ac{mimo} \ac{mmwave} communication system, wherein the transmission from a source (S) to a destination (D) is assisted by an \ac{ris} (I) as shown in Fig. \ref{fig:system_model}. The \ac{ris} I is controlled by D, and, with reference to a cellular system, this model corresponds to an uplink transmission scenario, where S is the user equipment and D is the base station. 

S and D are equipped with \acp{ula} with $N_{\rm S}$ and $N_{\rm D}$ antennas, respectively. The \ac{ris} has $N_{\rm I}$ passive reflective elements equally spaced along a line. We assume that the direct S-D channel is blocked, while connectivity between S and D is maintained through I.

We denote with $\bm{G} \in \mathbb{C}^{N_{\rm I} \times N_{\rm S}}$ and $\bm{H} \in \mathbb{C}^{N_{\rm D} \times N_{\rm I}}$ the narrowband S-I and I-D channel matrices, respectively. We adopt a block-fading channel model, thus both $\bm{G}$ and $\bm{H}$ remain constant during the channel coherence time.  

In the \ac{mmwave} band, channels have only a few relevant paths, thus we can use a geometric model for their description. Let $d$ be the spacing between both the \ac{ris} elements and the antennas at S and D; then, the array response column vector for an angle of arrival or departure $\beta$ is
\begin{align}
    [\bm{\alpha}(\beta)]_n &= e^{j 2\pi \frac{d}{\lambda} (n-1) \sin \beta}, \quad \quad  n = 1,\ldots,N_S.
\end{align}
For channel $\bm{G}$, let $[\bm{\rho}_{G}]_l$ be the $l$th propagation path gain, $L_{G}$ be the number of paths, vectors $\bm{\theta}_{G} \in [-\pi, \pi]$ and $\bm{\eta}_{G} \in [-\frac{\pi}{2}, \frac{\pi}{2}]$ be the \ac{aod} and the \ac{aoa} of the channel, respectively. %, $\bm{\alpha}([\bm{\theta}_{G}]_l)$ and $\bm{\alpha}([\bm{\eta}_{G}]_l)$ be the array response vectors as a function of $[\bm{\theta}_{G}]_l$ and $[\bm{\eta}_{G}]_l$. 
Here, we assume that both S and D are on the same side of the \ac{ris}, while the antennas of D can have any orientation on the plane. By following the geometric channel model, channel $\bm{G}$ with $L_G$ paths can be written as
\begin{align}
    \bm{G} &= \sum_{l=1}^{L_{G}}[\bm{\rho}_{G}]_l \bm{\alpha}([\bm{\eta}_{G}]_l)\bm{\alpha}^H([\bm{\theta}_{G}]_l) \\
           &= \bm{A}(\bm{\eta}_{G}) \diag(\bm{\rho}_{G}) \bm{A}^H(\bm{\theta}_{G}),
\label{eq:G} 
\end{align}
where $^H$ is the Hermitian operator and the array response matrices, i.e., $\bm{A}(\bm{\theta}_{G})$ and $\bm{A}(\bm{\eta}_{G})$, are defined as
\begin{align}
    \bm{A}(\bm{\theta}_{G}) &= \left[  \bm{\alpha}([\bm{\theta}_{G}]_1), ..., \bm{\alpha}([\bm{\theta}_{G}]_{L_{G}}) \right], \\
    \bm{A}(\bm{\eta}_{G}) &= \left[  \bm{\alpha}([\bm{\eta}_{G}]_1), ..., \bm{\alpha}([\bm{\eta}_{G}]_{L_{G}}) \right].
\end{align}
%where $\bm{\eta}_{G} \in [-\frac{\pi}{2}, \frac{\pi}{2}]$ and $\bm{\theta}_{H} \in [-\frac{\pi}{2}, \frac{\pi}{2}\pi]$ are the \ac{aoa} and the \ac{aod} of the channel, respectively, 
Similarly, channel $\bm{H}$ is modeled as 
\begin{align}
    \bm{H} &= \sum_{l=1}^{L_{H}}{[\bm{\rho}_{H}]_l \bm{\alpha}([\bm{\eta}_{H}]_l)\bm{\alpha}^H([\bm{\theta}_{H}]_l)}, \\
           &= \bm{A}(\bm{\eta}_{H}) \diag(\bm{\rho}_{H}) \bm{A}^H(\bm{\theta}_{H}),
\label{eq:H}        
\end{align}
where $[\bm{\rho}_{H}]_l$ is the $l$th propagation path gain, $L_{H}$ is the number of paths, $\bm{\theta}_{H} \in [-\frac{\pi}{2}, \frac{\pi}{2}]$ and $\bm{\eta}_{H} \in [-\frac{2}{6}\pi, \frac{2}{6}\pi]$ are the \ac{aod} and the \ac{aoa} of the channel, respectively. %, and $\bm{\alpha}([\bm{\eta}_{H}]_l)$ and $\bm{\alpha}([\bm{\theta}_{H}]_l)$ are the array response vectors. 
Here we assumed that the field of view of D is of $30^{\circ}$.

By applying (\ref{eq:G}) and (\ref{eq:H}), and considering the configuration of I, the complete end-to-end uplink channel between S and D is expressed as 
\begin{equation}
    \bm{Q} = \bm{H} \bm{\Phi} \bm{G},
\end{equation}
where $\bm{\Phi} \in \mathbb{C}^{N_{\rm I} \times N_{\rm I}}$ is the phase control matrix at I with unit-modulus elements on the diagonal, i.e., $[\bm{\Phi}]_{k,k}=e^{j \phi_k}$, with $\phi_k \in [0, 2\pi)$ and $k = 1,..., N_I$.

\section{Channel Estimation Protocol}
\label{FIC}

Our aim is to optimize $\bm{\Phi}$ to maximize the resulting achievable rate 
\begin{equation}
C(\bm{Q}) = \log_2 \det \left( \bm{I} + \frac{1}{\sigma^2}  \bm{Q}^H \bm{Q} \right),
\label{eq:cap}
\end{equation}
where $\sigma^2$ is the noise power.
We assume that channels $\bm{H}$ and $\bm{G}$ are not known to D. However, instead of first estimating the channels $\bm{H}$ and $\bm{G}$ (or, equivalently, the parameters $\bm{\eta}_H$, $\bm{\theta}_H$, $\bm{\rho}_H$, $\bm{\eta}_G$, $\bm{\theta}_G$, and $\bm{\rho}_G$) and then maximize the achievable rate, we directly seek the optimal \ac{ris} configuration by an iterative adaptive procedure, denoted \ac{fic}, where we leverage the fact that D has the control of I, thus can choose the configuration to be explored according to the performance obtained from previous configurations.

In particular, first S transmits pilot signals that let D estimate the resulting channel $\bm{Q}$ for a set of $L_1$ configurations 
\begin{equation}
\mathcal S_1 = \{\bm{\Phi}_{1,1}, \ldots, \bm{\Phi}_{1,L_1}\},
\end{equation}
and the resulting uplink channel estimates are $\hat{\bm{Q}}_{1,1}, \ldots, \hat{\bm{Q}}_{1,L_1}$. Based on these estimates, and on the computation of the resulting achievable rate $C(\hat{\bm{Q}}_{1,\ell})$, $\ell=1, \ldots, L_1$, D selects a new set of configurations, then S transmits pilots and D obtains the channel estimates for the new configurations. 

In general, at iteration $i>1$, S transmits pilots in correspondence of $L_i$ \ac{ris} configurations in the set
\begin{equation}
\mathcal S_i = \{\bm{\Phi}_{i,1}, \ldots, \bm{\Phi}_{i,L_i}\},
\end{equation}
and the resulting channel estimates are $\hat{\bm{Q}}_{i,1}, \ldots, \hat{\bm{Q}}_{i,L_i}$. D also computes the resulting achievable rate $C(\hat{\bm{Q}}_{i,\ell})$, $\ell=1, \ldots, L_i$. The \ac{fic} algorithm is stopped when a maximum number $I$ of iterations is reached. The choice of the configuration sets $\mathcal S_i$ is detailed in the following. 

%The \ac{fic} algorithm is stopped when a maximum number $I$ of iterations is reached.

The idea of the \ac{fic} algorithm is to progressively improve the precision of the selected configuration to maximize the resulting achievable rate. Moreover, we leverage the fact that \ac{mmwave} channels are characterized by a small number of paths. Note that we do not explicitly estimate all the channel parameters, e.g., we do not estimate the angles at the devices and the channel gains.

We first describe the \ac{fic} algorithm when a single path is present ($L_G = L_H$) and then we consider the general case with multiple paths.

\subsection{\ac{ris} Configuration With A Single-Path Channel}
\label{singlepathcase}

First, we observe that for single-path channels ($L_G = L_H = 1$), the \ac{ris} configuration maximizing the achievable rate aligns the steering vectors of the channels, i.e.,
\begin{equation}
    \bar{ \bm{\Phi}} = \diag\{e^{j\bar{\phi}_k}\} = \arg \max_{\Phi} ||\bm{A}^H(\bm{\theta}_{H})  \bm{\Phi} \bm{A}(\bm{\eta}_{G})||^{2},
\label{phidiag}
\end{equation}
where, for $k = 1,\ldots, N_I$,
\begin{equation}
    \bar{\phi}_k = 2\pi \frac{d}{\lambda} (\sin \bar{\theta} - \sin \bar{\eta})(k-1),
\end{equation}
with $\bar{\theta} = \theta_H$ and $\bar{\eta} = \eta_G$. We recall that the \ac{aoa} and \ac{aod} at the \ac{ris} are in the range $[-\frac{\pi}{2}, \frac{\pi}{2}]$ with respect to the firing direction of the linear element array.

Therefore, we design the \ac{ris} configurations $\bm{\Phi}_{1,\ell}$ as in (\ref{phidiag}), with a suitable choice of the angles $\bar{\theta}$ and $\bar{\eta}$.

At the first iteration ($i=1$) of the \ac{fic} algorithm, a regularly-spaced 2D grid of $L_1$ angle couples is designed inside a square with side $[-\frac{\pi}{2}, \frac{\pi}{2}]$. In particular, the phases of the \ac{ris} configuration are $[\bm{\Phi}_{1,\ell}]_{k,k}=e^{j \phi^{(1, \ell)}_k}$, with (for $i=1$ and $\ell=1,\ldots, L_1$)
\begin{equation}
\phi^{(i, \ell)}_k = 2\pi \frac{d}{\lambda}  (\sin \theta^{(i, \ell)} - \sin \eta^{(i, \ell)}) \left(k-1 \right) 
\label{phii}
\end{equation}
and 
\begin{equation}
\theta^{(1, \ell)} = \frac{\pi}{\sqrt{L_1}}\left[ \left\lfloor \frac{\ell -1}{\sqrt{L_1}} \right\rfloor  - \frac{\sqrt{L_{1}}}{2}\right] + \frac{\pi}{2\sqrt{L_1}},
\label{theta}
\end{equation}
\begin{equation}
\eta^{(1, \ell)} = \frac{\pi}{\sqrt{L_1}} \left[ \mod(\ell-1, \sqrt{L_1})   - \frac{\sqrt{L_{1}}}{2}\right] + \frac{\pi}{2\sqrt{L_1}},
\label{eta}
\end{equation}
where $\lfloor x \rfloor$ is the integer floor of $x$, and $\mod(x, L)$ is the reminder of $x/L$. 

The index of the \ac{ris} configuration $\bar{\bm{\Phi}}^{(1)}=\bm{\Phi}_{1,\bar{\ell}}$ providing the largest achievable rate is found as
\begin{equation}
\bar{\ell} = \arg \max_{\ell} C(\hat{\bm{Q}}_{1,\ell})
%\bar{\bm{\Phi}}^{(1)} = \arg \max_{\Phi \in \mathcal S_1} C(\Phi),
\end{equation}
and is used to design a new grid and start another iteration of the algorithm. 

In general, at iteration $i>1$, defining $\gamma_i = \sqrt{\prod_{k=1}^i L_k}$, the regularly-spaced 2D grid of $L_i$ cells is designed inside a square with side $\pi/\gamma_i$, and the phases of the \ac{ris} configuration $[\bm{\Phi}_{i,\ell}]_{k,k}=e^{j \phi^{(i, \ell)}_k}$ are given by \eqref{phii} and 
\begin{equation}
\theta^{(i, \ell)} = \frac{\pi}{\gamma_i} \left[ \left\lfloor \frac{\ell -1}{\sqrt{L_i}} \right\rfloor  - \frac{\sqrt{L_i}}{2}\right] + \frac{\pi}{2\gamma_i} + \bar{\theta}^{(i-1)},
\end{equation}
\begin{equation}
\eta^{(i, \ell)} = \frac{\pi}{\gamma_i} \left[ \mod(\ell-1, \sqrt{L_i})   - \frac{\sqrt{L_i}}{2}\right] + \frac{\pi}{2\gamma_i} + \bar{\eta}^{(i-1)},
\end{equation}
where $\bar{\theta}^{(i-1)}$ and $\bar{\eta}^{(i-1)}$ are the angles of the RIS configuration $\bar{\bm{\Phi}}^{(i-1, \bar{\ell})}$ selected at the previous iteration with 
\begin{equation}
\bar{\ell} = \arg \max_{\ell \in \{ 1, ..., L_{i-1}\}} C(\hat{\bm{Q}}_{i-1,\ell}).
\end{equation} 
When $i=I$, the procedure stops providing the  \ac{ris} configuration  $\bar{\bm{\Phi}}^{(I, \bar{\ell})}$.

\paragraph*{Multiple Starting Points} The iterative procedure in general will find a local maximum of the achievable rate, as the achievable rate is not a convex function of the \ac{aoa} and \ac{aod}. To improve the performance of \ac{fic}, multiple starting points can be considered. In this case, $P$ \ac{ris} configurations are selected from those explored at the first iteration, and the iterative procedure is applied on all the $P$ selected angle couples to find $P$ local maxima. Then, the configuration yielding the maximum achievable rate is selected as the final solution. Note that this increases the duration of the \ac{fic} algorithm while bringing it closer to the optimal \ac{ris} configuration. 
 
\subsection{\ac{ris} Configuration With Multi-Path Channel}
\label{risMulti}

\begin{figure}
 \centering
 \includegraphics[width=\columnwidth]{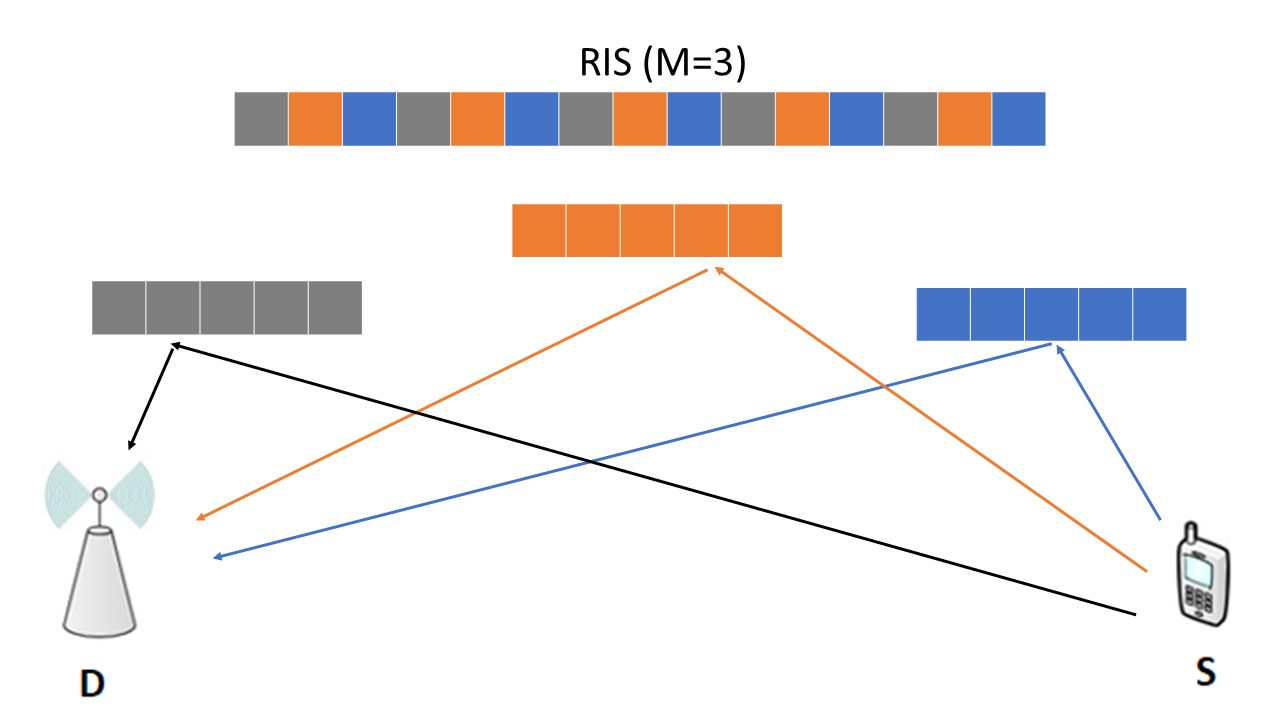}
  \caption{Example of multi-path scenario with $M=3$.}
  \label{fig:multipath}
\end{figure}

When multiple paths are present, we partition the \ac{ris} elements into $M$ sets of $N_I/M$ elements each, where set $m$ will have phases aligned to the $m$-th path, between S and I with the $m$-th path between I and D. Since one \ac{aoa} at the \ac{ris} will be matched to a single \ac{aod} from the \ac{ris}, we will consider $M= \min\{L_G, L_H\}$ configurations, as shown in Fig.~\ref{fig:multipath} for $M=3$.

Therefore, we repeat the algorithm of Section~\ref{singlepathcase} to find the configuration of the various \acp{ris}. Since \ac{ris} elements cannot be switched off, we propose the following procedure, operating in $M$ steps, where in each step we select the configuration for $N_I/M$ \ac{ris} elements.  

In the first step, we configure all the $N_I$ \ac{ris} elements using the iterative procedure of Section~\ref{singlepathcase}. Then, we fix the obtained configuration for $N_I/M$ elements, starting from the first and regularly spaced by $M-1$ elements. 

In step $m>1$, we optimize the 
\begin{equation}
N_I - (m-1)\frac{N_1}{M}
\end{equation}
elements not yet fixed, and then we fix the obtained configuration for $N_I/M$ elements, starting from the $m$-th and regularly spaced by $M-1$ elements. When configuring the set $m$ of the \ac{ris} elements, we compute the achievable rate \eqref{eq:cap} by considering the already set values of the phases for the $m-1$ sets already optimized. 

Note that we do not put additional constraints on the configuration selection procedure. Thus, it may happen that more than one \ac{ris} subset is configured for the same angles, thus serving the same path. This occurs when the achievable rate gain obtained by a stronger reflection is higher than that associated with the larger multiplexing provided by reflecting an additional path. Still, our algorithm performs this tradeoff implicitly, without the need for further elaboration.

Lastly, note that although the \ac{ris} configurations are tuned to specific \ac{aoa}s and \ac{aod}s of the \ac{ris}, they also contribute to the reflection of other paths, thus each sub-block of the \ac{ris} is in practice not strictly associated to a single path but contributes to the whole resulting achievable rate in a more elaborate way.

\subsection{Estimation Time}

Assuming that the number of pilot symbols for each \ac{ris} configuration is fixed and their transmission takes time $T_0$, the total time spent by the \ac{fic} algorithm is 
\begin{equation}
T_{\rm FIC} = T_0 M \left(L_1 +P\sum_{i=2}^I L_i\right),
\end{equation}
since to configure each of the $M$ sub-array of the \ac{ris} we need $I$ iterations, each requiring the exploration of $L_i$ \ac{ris} configurations. Moreover, we have taken into account the possibility of using $P$ starting points.

\section{Numerical results}
\label{sec:numres} 

In this section, we assess the performance of the proposed \ac{fic} algorithm. We consider a single-user cellular uplink transmission system assisted by a \ac{ris} and an urban micro-cell (UMi) scenario for a \ac{mmwave} transmission at the carrier frequency of $28$~GHz, considering the geometry-based stochastic channel model used in 3GPP  \cite{hur2016}. The \ac{snr} is set at $-15$~dB. 

We assume that both S-I and I-D links exhibit a \ac{los} condition remarking that the direct link between S and D is blocked.  S and D are equipped with an \ac{ula} array of $N_S=2$ and $N_D=4$ antennas, respectively. For the \ac{ris}, we consider $N_I=120$ elements along a line %;  two values of element/antenna spacings $d$ are considered, namely $d=\lambda/2$ and 
spaced $d=\lambda/2$, where $\lambda$ is the wavelength at the carrier frequency. We assume $T_0=1$, thus $T$ counts the number of channel estimates.

For comparison purposes, we consider a scheme wherein the channel is estimated for a set of \ac{ris} configurations and select the one providing the highest achievable rate. This corresponds to stopping the interactive procedure on the first iteration $I=1$ while considering denser grids (i.e., larger values of $L_1$). This scheme is denoted by baseline (BAS) configuration.

Performance is evaluated in terms of the average normalized achievable rate loss (averaged over the channel and noise realizations)
\begin{equation}
\epsilon = {\mathbb E}\left[\dfrac{C_{opt}-\hat{C}}{C_{opt}}\right],
\end{equation}
where $C_{opt}$ is the achievable rate obtained with the optimal \ac{ris} configuration, whereas $\hat{C}$ is the achievable rate obtained with the \ac{ris} configuration of the \ac{fic} algorithm (or BAS). % Note that according to what stated at the end of Section~\ref{risMulti}, 
For multi-path channels, $C_{opt}$ must be determined through an exhaustive search.% on a 2D grid of $L_1$ \ac{ris} configurations, with $L_1$ large enough to have a fine grid. 

\begin{figure}
 \centering
\includegraphics[width=0.9\columnwidth]{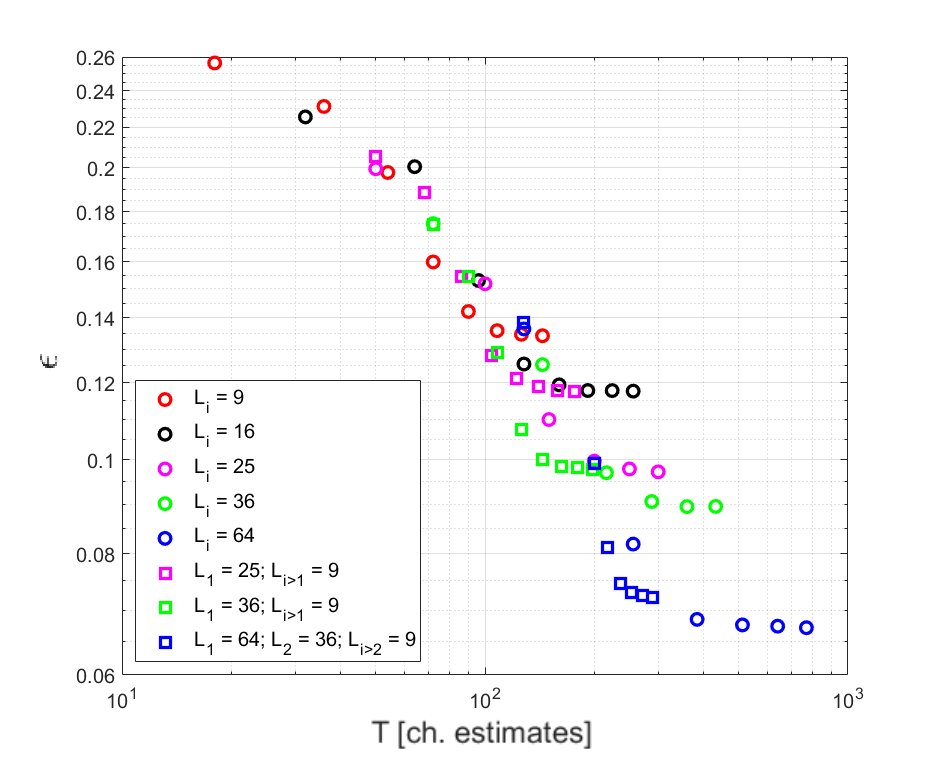}
  \caption{Average normalized achievable rate loss varying $L_i$ with $L_G = L_H = 2$ paths and element/antenna spacing $d=\frac{\lambda}{2}$.}
  \label{fig:2paths}
\end{figure}

\begin{figure}
 \centering
\includegraphics[width=\columnwidth]{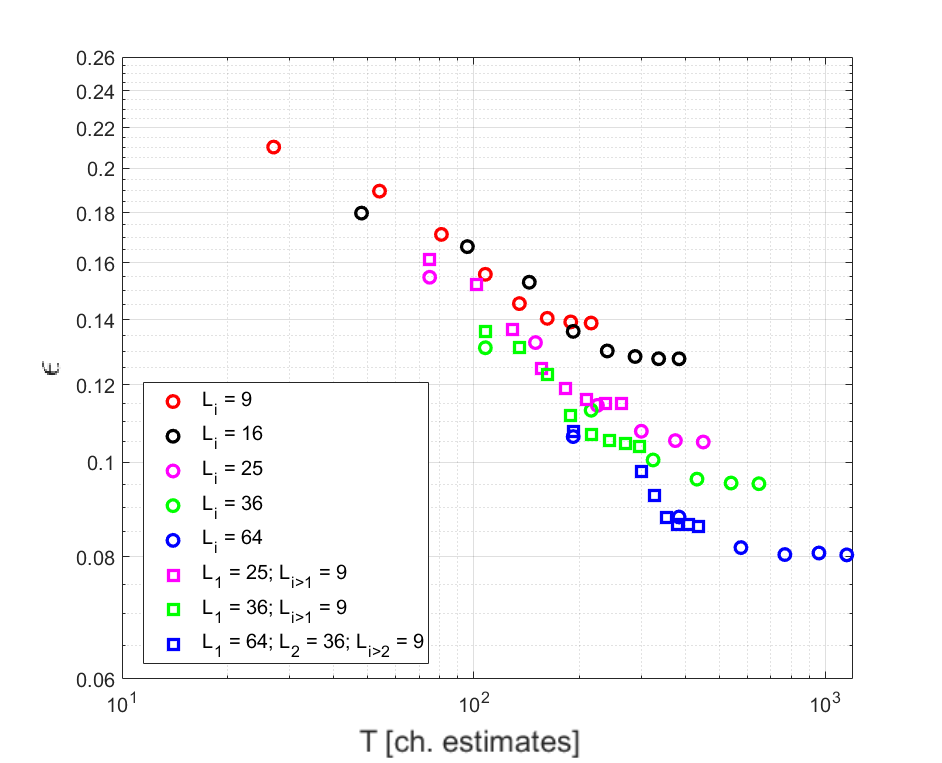}
  \caption{Average normalized achievable rate loss varying $L_i$ with $L_G = L_H = 3$ paths and element/antenna spacing $d=\frac{\lambda}{2}$.}
  \label{fig:3paths_1}
\end{figure}

Figs. \ref{fig:2paths} and \ref{fig:3paths_1} show the average normalized achievable rate loss $\epsilon$ as a function of the total number of channel estimates $T$ for an element/antenna spacing $d=\lambda/2$. Two values of the number of channel paths are considered, namely, $L_G=L_H=2$ and  $L_G=L_H=3$, and for the choice of the configuration the  \ac{ris} is split into $M=2$ and 3 sub-blocks, respectively. On the grid choice for \ac{fic}, we consider two cases: a) $L_i$ is constant at all iterations, and b) $L_i$ changes at the various iterations. In particular, for case a) we considered $L_i = 9$, 16, 25, 36, and 64, while for case b) we considered the following alternatives: 1) $L_1=25$, $L_i = 9$ for $i>1$, 2) $L_1=36$, $L_i = 9$ for $i>1$, and 3) $L_1=64$, $L_2=36$, $L_i = 9$ for $i>2$. For the BAS, we consider $L_1=9$, 16, 25, 36, 64, 100, 225, and 400.

For \ac{fic}, the results for different values of $T$ are obtained by varying the total number of iterations $I$. %, while for BAS we only change the value of $L_1$. 
As expected, a higher number of iterations yields a lower achievable rate loss, at the cost of a higher channel estimation overhead, and we note that within $10^3$ estimations the achievable rate loss is less than 10\%. %When compared to the BAS configuration, we note that \ac{fic} significantly improves the quality of the selected configuration, with a sharp decrease in achievable rate loss. 
We also note that, for a higher value of $L_i$, a lower %average normalized achievable rate loss 
$\epsilon$ is typically obtained. Moreover, changing the number of explored \ac{ris} configurations across the iterations yields a further performance improvement. Lastly, when comparing Figs. \ref{fig:2paths} and \ref{fig:3paths_1}, we note that when more paths are present a longer time $T$ is required, and the finer the grid, the higher the minimum value of the  achievable rate loss: indeed, the function relating the angles at I (thus the \ac{ris} configuration) to the achievable rate is more complex, and it is harder to find the global maximum.  Furthermore, it can be seen that for many of the considered grids $\epsilon$ tends to saturate without further reducing the achievable rate loss. This is due to the fact that the algorithm locks in a local maximum.

\begin{figure}
 \centering
  \includegraphics[width=\columnwidth]{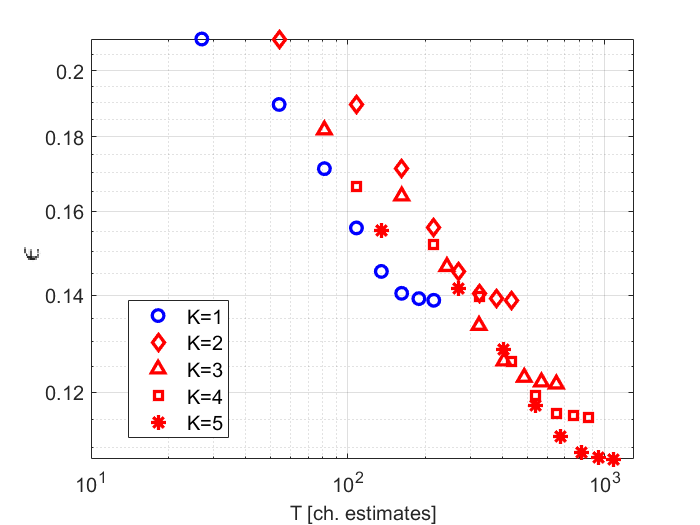}
  \caption{Average normalized achievable rate loss varying the number of channel estimates per iteration $K = 1$, 2, 3, 4, and 5 with $L_i=9$ $L_G = L_H = 3$ paths and element/antenna spacing $d=\frac{\lambda}{2}$}
  \label{fig:3paths_snrvar}
\end{figure}

\paragraph*{Impact of the Number of Channel Estimates} Until now we have considered a single channel estimate at each iteration of the \ac{fic} algorithm: such estimates are affected by the noise, which has also an impact on the estimate of the achievable rate and thus the evolution of the \ac{ris} configuration search in the \ac{fic} algorithm. Performing $K>1$ channel estimates and then averaging the estimates before computing the achievable rate would provide a better result, at the cost of a longer estimation time, since, still assuming a unitary time for a single estimate, we have $T_0 = K$. Fig.~\ref{fig:3paths_snrvar} shows %the average normalized achievable rate loss 
$\epsilon$ as a function of $T$ for a number of estimates per iteration $K = 1$, 2, 3, 4, and 5. %We note that t
The error floor of $\epsilon$ is reduced as $K$ increases, thus if low achievable rate losses are targeted it is advantageous to increase the number of channel estimates per iteration. However, whenever a value of $\epsilon$ is reachable for multiple values of $K$, it is convenient to choose the minimum $K$ to minimize the estimation time $T$.

\begin{figure}
 \centering
\includegraphics[width=\columnwidth]{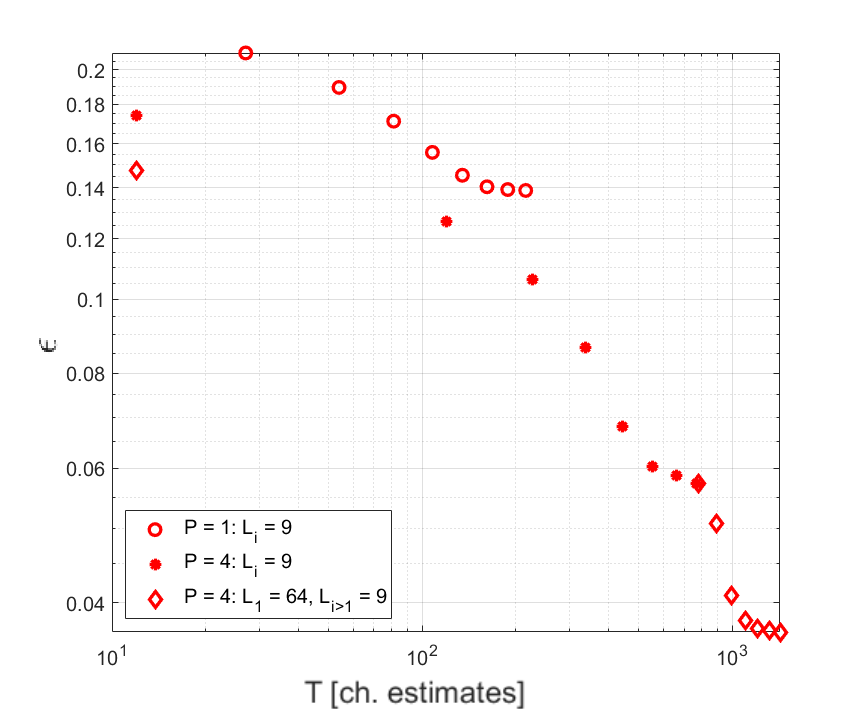}
  \caption{Average normalized achievable rate loss with $P=1$ and 4 starting points, $L_i=9$, $L_G = L_H = 3$ paths, and element/antenna spacing $d=\frac{\lambda}{2}$. }%\hl{LEGENDA: $P=1$ e $P=2$}}
  \label{fig:multistart}
\end{figure}

\paragraph*{Multiple Starting Points} We have also tested the multiple starting points approach, and Fig.~\ref{fig:multistart} shows $\epsilon$ as a function of $T$ for $P=1$ and 4 starting points, $L_G = L_H = 3$ paths, and $d=\frac{\lambda}{2}$. In particular, we consider $L_{i}=9$, $i=1,.., I$, and, in addition to this, for $P=4$ also the case $L_{1}=64$, $L_{i>1}=9$. The $P$ starting points provide the highest achievable rate among the $P$ configurations tested at the first iteration. We note an advantage of the multiple starting point approach at the cost of almost four times the number of channel estimates.

\begin{figure}
 \centering
  \includegraphics[width=\columnwidth]{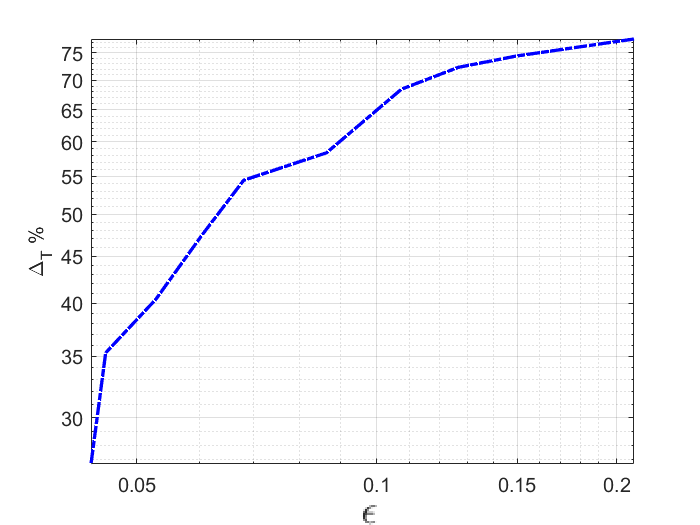}
  \caption{Percentage reduction of $T_{FIC}$ with respect to $T_{BAS}$ for a target $\epsilon$, with $L_G = L_H = 3$ paths and element/antenna spacing $d=\frac{\lambda}{2}$.}
  \label{fig:3paths_comparison}
\end{figure}

\paragraph*{Comparison With Respect to BAS}
Fig. \ref{fig:3paths_comparison} shows the percentage reduction of $T_{FIC}$ with respect to $T_{BAS}$, named $\Delta_T$ \%, to obtain a target $\epsilon$ for $d=\lambda/2$ and $L_G=L_H=3$.  
When compared to the BAS configuration, we note that \ac{fic} significantly improves the quality of the selected configuration, with a sharp decrease in the achievable rate loss. This turns in a decrease of at least about 30\% in the number of channel estimates required to reach a target $\epsilon$.

\section{Conclusions}
\label{sec:conc}
 
In this paper, we have proposed a novel technique to find the configuration of a \ac{ris} for the uplink of a cellular system operating at \ac{mmwave} frequencies. %The procedure is iterative and adaptive, as the explored configurations are chosen according to the achievable rate obtained on the channels resulting from the previously explored configurations. Moreover, the procedure exploits the presence of a few paths in the \ac{mmwave} channel by splitting the \ac{ris} into multiple sub-\acp{ris}, each optimized for a single path. 
Numerical results have compared the performance of the proposed approach with a baseline solution where a fixed set of configurations is explored and the configuration providing the highest achievable rate is selected. 
We conclude that the proposed procedure is effective and shows a much faster convergence to a close-to-optimal configuration. We have also explored the impact of % antenna spacing and 
the number of channel estimates per iteration and the multiple-starting-point approach for further optimization of the proposed scheme.

\end{document}